\documentclass[twocolumn,notitlepage,superscriptaddress]{revtex4-1} 
\usepackage[colorlinks=true,urlcolor=blue,linkcolor=blue,citecolor=blue]{hyperref}
\usepackage{latexsym}
\usepackage{graphicx}
\usepackage{subfigure} 
\usepackage{placeins}
\usepackage{amsmath}
\usepackage{color}
\usepackage{amsfonts}
\usepackage{mathrsfs}

\usepackage[most]{tcolorbox} 
\newtcolorbox{techdefinition}{
    enhanced, 
    frame style={left color=blue!70!black, right color=blue!30!black}, 
    colback=blue!3!white, 
    colframe=white, 
    boxrule=0pt, 
    arc=3pt, 
    fonttitle=\bfseries\sffamily\large, 
    coltitle=blue!70!black, 
    attach title to upper, 
    borderline vertical={1pt}{0pt}{blue!50!white}, 
    before skip=1.5\baselineskip, 
    after skip=1.5\baselineskip, 
}

\begin{document}

\title{Coincidence detection techniques for direct measurement of many-body correlations in strongly correlated electron systems} 

\author{Yuehua Su}
\email{suyh@ytu.edu.cn}
\affiliation{ Department of Physics, Yantai University, Yantai 264005, People's Republic of China }

\author{Guoya Zhang}
\affiliation{ Department of Physics, Yantai University, Yantai 264005, People's Republic of China }

\author{Chao Zhang}
\affiliation{ Department of Physics, Yantai University, Yantai 264005, People's Republic of China }

\author{Dezhong Cao}
\affiliation{ Department of Physics, Yantai University, Yantai 264005, People's Republic of China }

\begin{abstract}

Research on strongly correlated electron systems faces a fundamental challenge due to the complex nature of intrinsic many-body correlations. A key strategy to address this challenge lies in advancing experimental methods that can directly probe and elucidate the underlying many-body correlations. In this perspective article, we discuss the theoretically proposed coincidence detection techniques, which are designed to directly measure two-body correlations in various particle-particle and particle-hole channels, with momentum, energy, and/or spatial resolution. We also explore the prospects of these coincidence detection techniques for future theoretical and experimental developments. The successful implementation and refinement of these coincidence detection techniques promise to deliver powerful new approaches for unraveling long-standing puzzles in strongly correlated electron systems, such as the enigmatic mechanism of unconventional superconductivity and the long-sought quantum spin liquids. Furthermore, these coincidence detection techniques will offer powerful new methods to investigate novel phenomena like itinerant magnetism and electronic nematicity in quantum materials.

\end{abstract}


\maketitle

\section{Introduction} \label{sec1}

Strongly correlated electron systems represent a major focus of modern condensed matter physics due to their novel physical properties. Systems such as unconventional superconductors \citep{AndersonScience1987, PALeeRMP2006, ChenXHNAR2014, StewartFeSCRMP2011, Stewart1984, StewartNFLRMP2001} and quantum magnets \citep{BalentsNature2010, ZhouYiRMP2017, ChamorroCR2021} exhibit a range of unconventional and anomalous phenomena, including non-BCS superconductivity, pseudogap physics, strange metal behavior, electronic nematicity, quantum spin liquids, and quantum phase transitions. These emergent properties are governed by intrinsic many-body correlations that can not be adequately described within well established theoretical frameworks like the perturbative Fermi liquid theory or the Ginzburg-Landau-Wilson theory of phase transitions. The development of new theoretical and experimental methods to study these correlation-driven phenomena remains a significant challenge. In this perspective article, we focus on recent progress and ongoing efforts to develop new experimental techniques within this research field. 

A variety of experimental techniques have been developed to investigate the superconducting states of unconventional superconductors. Zero electric resistance and the diamagnetic Meissner effect are two key signatures of the superconducting phase, while specific heat measurements establish its macroscopic phase transition nature. Evidence for Cooper-pair-mediated superconductivity comes from flux quantization experiments \citep{GoughNature1987}. The superfluid density is accessible via measurements of the magnetic penetration depth \citep{PanagopoulosXiangPRL1998}, and the low-energy optical conductivity reveals the electrodynamic response of the superfluid condensate and the corresponding charge-carrier excitations \citep{BasovRMP2005}. For pairing symmetry, angle-resolved photoemission spectroscopy (ARPES) \citep{ShenRMP2003} and scanning tunneling spectroscopy (STS) \citep{FischerRMP2007} offer momentum and spatial resolutions, respectively. Further insights are provided by nuclear magnetic resonance (NMR) \citep{Asayama1991} and electronic Raman spectroscopy \citep{DevereauxRMP2007}. Phase-sensitive techniques have been especially critical in identifying the pairing symmetry of high-Tc cuprate superconductors \citep{TsueiRMP2000}. Together, these methods enable a multifaceted investigation of superconductivity across different channels and scales. However, a fundamental limitation remains that none directly probes the intrinsic two-body correlations of Cooper pairs. Consequently, they can not unequivocally determine the microscopic pairing mechanism responsible for high-Tc superconductivity. 

Beyond the mystery of high-Tc superconductivity itself, the parent normal states in unconventional superconductors exhibit equally esoteric properties. In cuprate superconductors, a pseudogap phase manifests in transport, thermodynamic, and magnetic channels \citep{LiTao2021, LuoSuXiang2008, ProustTaillefer2019}, while the strange metal state is characterized by an enigmatic linear-$T$ resistivity \citep{KeimerNature2015, LiTao2021, ProustTaillefer2019, VarmaXYRMP2020, Zaanen2019}. Similarly, electronic nematicity is ubiquitous in iron-based superconductors, with ARPES measurements revealing a d-wave symmetric feature within the nematic phase \citep{YiPANS2011, SuLi2015}. Heavy-fermion systems, in turn, display non-Fermi liquid behaviors and diverse quantum critical phenomena, as observed in specific heat, magnetic susceptibility, and resistivity measurements \citep{StewartNFLRMP2001}. These phenomena collectively point to strong many-body correlations that remain elusive to direct experimental probes. 

A particularly intriguing class of quantum states in strongly correlated electron systems is the theoretically proposed quantum spin liquids \citep{AndersonScience1987}. The investigation of these states remains a prominent and active frontier in condensed matter physics \citep{BalentsNature2010, ZhouYiRMP2017, SavaryBalents2017, ChamorroCR2021}. Experimentally, techniques such as muon spin spectroscopy, inelastic neutron scattering (INS), NMR, and thermal transport measurements are employed to study their magnetic, thermodynamic, and transport properties \citep{SavaryBalents2017}. Theoretically, models based on spin fractionalization predict the emergence of fractionalized excitations in these systems \citep{PALeeRMP2006, Auerbachbook, KitaevAP2006, GaoChenCPB2020}. However, such fractionalized excitations do not directly couple to conventional experimental probes, thereby presenting a major challenge for their direct detection and characterization. 

As outlined in the above brief review, a wide array of experimental techniques has been employed to investigate strongly correlated electron systems. Despite the substantial body of experimental data thus accumulated, research in this field appears to be approaching a bottleneck. This impasse may be attributed to two principal factors. First, the essential intrinsic signatures have yet to be systematically distilled from the vast and complex experimental data, hindering the development of sound phenomenological theories. Second, the existing experimental approaches may be fundamentally limited in their ability to probe the intrinsic many-body correlations and their physical manifestations. Breaking through this impasse and unraveling the foundational correlation-driven physics will likely require the development of new experimental theories and advanced technologies. 

\begin{widetext}

\begin{figure}[ht]
\includegraphics[width=1.0\columnwidth]{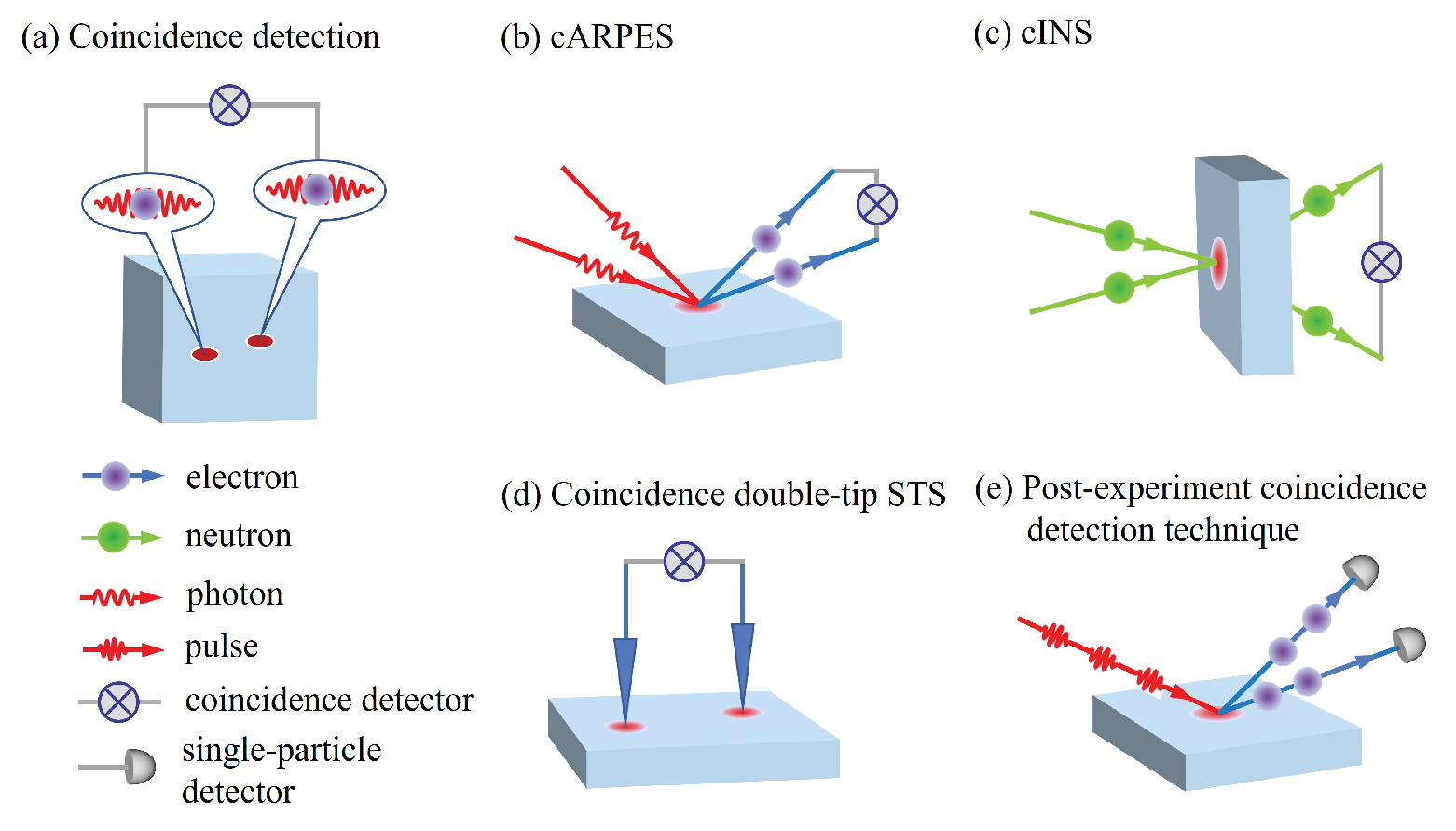}  
\caption{ Schematic illustration of coincidence detection. (a) Fundamental idea: Coincidence detection of two single-particle perturbation processes reveals dynamical two-body correlations. (b) By coincidence detection of two photoelectric processes,  coincidence angle-resolved photoemission spectroscopy (cARPES) captures the dynamical two-body correlations in the particle-particle channel. (c) By coincidence detection of two neutron-scattering processes, coincidence inelastic neutron scattering (cINS) uncovers dynamical two-spin correlations. (d) By coincidence detection of two quantum tunneling currents, coincidence double-tip scanning tunneling spectroscopy (STS) probes the spatially resolved dynamical two-body correlations. (e) Coincidence detection technique with post-experiment coincidence counting method. Other coincidence detection techniques, such as coincidence angle-resolved photoemission and inverse-photoemission spectroscopy (cARP/IPES), coincidence angle-resolved inverse-photoemission spectroscopy (cARIPES), and double photoemission (DPE), share a similar configuration to (a), but with different incident/scattered particle sources or detectors, for the dynamical two-body correlations in the different particle-particle or particle-hole channel. }
\label{fig1}
\end{figure}
\end{widetext}

The intrinsic $N$-body correlations in strongly correlated electron systems can be accessed via $N$-body response measurements. For example, the intrinsic two-body correlations can be directly probed via two-body response functions which are dominated by the second-order perturbation of the interaction between the target material and an external probe field. Guided by this principle, several coincidence detection techniques have recently been proposed for directly measuring two-body correlations in strongly correlated electron systems, as schematically illustrated in Fig.~\ref{fig1}. Notable examples include coincidence angle-resolved photoemission spectroscopy (cARPES), which is designed to access two-body correlations in the particle-particle channel, thereby offering a new pathway to elucidate the mechanism of unconventional superconductivity \citep{SuZhang2020, DevereauxPRB2023, CaoSuArXiv2024NJP}. Meanwhile, coincidence inelastic neutron scattering (cINS) has been proposed to probe two-spin correlations, opening avenues for studying exotic quantum spin liquids \citep{SucINS2021, CaoSuArXiv2024NJP}. More recently, a coincidence double-tip scanning tunneling spectroscopy (STS) method has also been introduced, enabling spatially resolved detection of dynamical two-body correlations \citep{SuArXiv2025}. Other proposed coincidence detection techniques include coincidence angle-resolved photoemission and inverse-photoemission spectroscopy (cARP/IPES) in the particle-hole channel for studying itinerant magnetism and electronic nematicity \citep{SuZhang2020}, coincidence angle-resolved inverse-photoemission spectroscopy (cARIPES) in the particle-particle channel for correlated electrons above the Fermi energy \citep{SuZhang2020}, and double photoemission (DPE) for exploring center-of-mass physics of two-body correlations \citep{BerakdarPRB1998, SuYHJPCM2024}. 
The fundamental principles underlying these coincidence detection techniques can be summarized as follows \citep{SuYHJPCM2024}: (1) Intrinsic two-body correlations are directly captured through two-body coincidence detection, and (2) the coincidence detection probability is governed by the second-order perturbation of the interaction between the target material and the external probe field.

In this perspective article, we will explore the conceptual foundations of these coincidence detection techniques, covering their key theoretical formulations and applications for probing two-body correlations in various specific channels. We will also address the challenges and prospects for their theoretical development and experimental implementation.

\section{Coincidence Detection: Fundamental Principles} \label{sec2}

We begin with the Hamiltonian of the combined system comprising the target electron system and the external probe field,
\begin{equation}
H = H_0 + V , \label{eqn2.1}
\end{equation}
where $H_0$ includes the individual Hamiltonians of the two subsystems, and $V$ describes the electron-probe interaction terms involving single-electron scattering processes. Let us introduce the time-evolution $S$-matrix defined as
\begin{equation}
S(t, t_i) = T_t \exp\left[ -\frac{i}{\hbar} \int_{t_i}^t d t_1 V_I(t_1) \right], \label{eqn2.2}
\end{equation}
where $T_t$ is a time ordering operator, and $V_I$ denotes the representation of $V$ in the interaction picture:
\begin{equation}
V_I(t) = e^{i H_0 t / \hbar} V e^{-i H_0 t / \hbar}. \label{eqn2.3}
\end{equation}
Here, $H_0$ is assumed to be time-independent.

Let the initial states of the combined system at time $t_i$ be described by ${| \Phi_I \rangle}$ with statistical probabilities $P_I$, where $I$ labels the state index. The first-order perturbation of the $S$-matrix, corresponding to single-electron scattering processes, is given by
\begin{equation}
S^{(1)}(t,t_i)=-\frac{i}{\hbar} \int_{t_i}^t d t_1 V_I(t_1) . \label{eqn2.4}
\end{equation}
The associated single-electron scattering probability reads
\begin{equation}
\Gamma^{(1)} = \sum_{I F} P_I \left| \langle \Phi_F | S^{(1)}(t_f, t_i) | \Phi_I \rangle \right|^2 , \label{eqn2.5}
\end{equation}
where ${| \Phi_F \rangle}$ are the final states at time $t_f$, labeled by index $F$.
Similarly, the second-order perturbation expansion takes the form
\begin{equation}
S^{(2)}(t,t_i)=\frac{1}{2}\left(-\frac{i}{\hbar}\right)^2 \iint_{t_i}^t d t_2 d t_1 T_t[V_I(t_2) V_I(t_1)] , \label{eqn2.6}
\end{equation}
which captures two-electron scattering processes via the time-ordered product $T_t[V_I(t_2) V_I(t_1)]$. The corresponding coincidence detection probability of two-electron scattering processes is defined by
\begin{equation}
\Gamma^{(2)} = \sum_{I F} P_I \left| \langle \Phi_F | S^{(2)}(t_f, t_i) | \Phi_I \rangle \right|^2 . \label{eqn2.7}
\end{equation}

While $\Gamma^{(1)}$ involves single-electron scattering processes and thus probes the dynamical single-electron physics of the target electron system, $\Gamma^{(2)}$ arises from two-electron scattering processes and captures the dynamical two-body correlations. This constitutes the conceptual foundation of coincidence detection, which uses the measurement of two-electron scattering processes to reveal the intrinsic two-body correlations in the target electron system. The fundamental principles underlying coincidence detection can be formally stated as follows \citep{SuYHJPCM2024}:
\begin{enumerate}
\item Intrinsic two-body correlations are directly captured through two-body coincidence detection.
\item The coincidence detection probability is governed by the second-order perturbation of the interaction between the target material and the external probe field.
\end{enumerate}
These principles provide a unified framework for the variety of coincidence detection techniques introduced in recent works \citep{SuZhang2020, SucINS2021, DevereauxPRB2023, CaoSuArXiv2024NJP, SuArXiv2025}.

Beyond the coincidence detection probability $\Gamma^{(2)}$, coincidence detection can also be formulated in terms of operator expectations. Consider an operator $O$ describing single-electron scattering processes. Its expectation value is given by
\begin{equation}
\overline{\langle O (t) \rangle} = \langle S(t_i, t) O_I(t) S(t, t_i) \rangle , \label{eqn2.8}
\end{equation}
where $\langle A \rangle \equiv \operatorname{Tr} [\rho_0 A]$ with $\rho_0 = \frac{1}{Z}\exp (-\beta H_0)$, and $S(t_i, t) = [S(t, t_i)]^\dag$. Introducing the contour-time ordered $S$-matrix,
\begin{equation}
S_c(t_i, t_i)=T_c [S(t_i,t) S(t, t_i)] , \label{eqn2.9}
\end{equation}
where $T_c$ is the contour-time ordering operator along the contour $t_i \rightarrow t \rightarrow t_i$, we may write $S_c$ explicitly as
\begin{equation}
S_c(t_i, t_i) = T_c \exp \left[ -\frac{i}{\hbar} \int_c d t_1 V_I(t_1) \right] , \label{eqn2.10}
\end{equation}
with the integration performed along the closed time contour. The expectation value of $O$ then becomes
\begin{equation}
\overline{\langle O (t) \rangle} = \langle T_c S_c(t_i, t_i) O_I(t) \rangle . \label{eqn2.11}
\end{equation}
For two operators $O_1$ and $O_2$, each describing single-electron scattering processes, the coincidence detection expectation value is defined by
\begin{equation}
\overline{\langle O_1 (t) O_2(t) \rangle} = \langle T_c S_c(t_i, t_i) O_{1 I} (t) O_{2 I}(t) \rangle . \label{eqn2.12}
\end{equation}
This expression provides the fundamental formulation for coincidence detection in the context of coincidence double-tip STS measurements \citep{SuArXiv2025}.

\section{Coincidence Detection Techniques} \label{sec3}

We will explore the coincidence detection techniques in this section, which are proposed based upon the fundamental principles summarized in Sec. \ref{sec2}.

\subsection{{cARPES}} \label{sec3.1}

\begin{techdefinition}
\noindent {\it Coincidence detection of two photoelectric processes, in which two incident photons emit two photoelectrons from the target material, captures two-body correlations of the correlated electrons in the particle-particle channel.} --- \text{cARPES}
\end{techdefinition}

In cARPES coincidence detection, as schematically illustrated in Fig.~\ref{fig1} (b), two incident photons induce two photoelectric processes. A coincidence detector records the simultaneous occurrence events of two photoelectric processes. The corresponding coincidence detection probability carries information on two-body correlations of the target electron system.

Suppose the two photoelectric processes are governed by two electron-photon interaction terms
\begin{eqnarray}
V_1 &=& g_1 d^{\dag}_{\mathbf{k}_1+\mathbf{q}_1 \sigma_1} c_{\mathbf{k}_1\sigma_1} a_{\mathbf{q}_1\lambda_1} , \notag \\
V_2 &=& g_2 d^{\dag}_{\mathbf{k}_2+\mathbf{q}_2 \sigma_2} c_{\mathbf{k}_2\sigma_2} a_{\mathbf{q}_2\lambda_2} , \label{eqn3.1.1}
\end{eqnarray}
where $a_{\mathbf{q} \lambda}$ annihilates an incident photon with momentum $\mathbf{q}$ and polarization $\lambda$, $c_{\mathbf{k} \sigma}$ annihilates a target electron with momentum $\mathbf{k}$ and spin $\sigma$, and $d^{\dag}_{\mathbf{k} \sigma}$ creates a photoelectron. The coupling constants $g_1 \equiv g(\mathbf{k}_1;\mathbf{q}_1\lambda_1)$ and $g_2 \equiv g(\mathbf{k}_2;\mathbf{q}_2\lambda_2)$ characterize the interaction strengths.

Let us introduce a two-body Bethe-Salpeter wave function in the particle-particle channel as \citep{GellmanLowBS1951,SalpeterBethe1951}
\begin{equation}
\phi^{(2)}_{\alpha\beta}\left( \mathbf{k}_1\sigma_1 t_1; \mathbf{k}_2\sigma_2 t_2 \right) = \langle \Psi_{\beta} \arrowvert T_t c_{\mathbf{k}_2 \sigma_2} \left(t_2\right) c_{\mathbf{k}_1 \sigma_1} \left(t_1\right) \arrowvert \Psi_{\alpha} \rangle , \label{eqn3.1.2}
\end{equation} 
where $\lvert \Psi_\alpha \rangle$ and $\lvert\Psi_\beta \rangle$ are eigenstates of the target electron system with energies $E_\alpha$ and $E_\beta$, respectively. 
In the limit $t_i \to -\infty$ and $t_f \to +\infty$, the coincidence detection probability for the simultaneous occurrence of two photoelectric processes  follows \citep{SuZhang2020}
\begin{equation}
\Gamma^{(2)} = \frac{\left( g_1 g_2 \right)^2 }{ \hbar^4} \frac{1}{Z} \sum_{\alpha\beta} e^{-\beta E_{\alpha}} \big\vert \phi^{(2)}_{\alpha\beta}\left(\mathbf{k}_1\sigma_1, \mathbf{k}_2\sigma_2; \Omega_c, \omega_r \right) \big\vert^2 . \label{eqn3.1.3}
\end{equation}
Here $\phi^{(2)}_{\alpha\beta}\left(\mathbf{k}_1\sigma_1, \mathbf{k}_2\sigma_2; \Omega_c, \omega_r \right)$ is the frequency-space Bethe-Salpeter wave function.  It is the Fourier transformation of  $\phi^{(2)}_{\alpha\beta}\left( \mathbf{k}_1\sigma_1, \mathbf{k}_2\sigma_2; t_c, t_r \right) = \phi^{(2)}_{\alpha\beta}\left( \mathbf{k}_1\sigma_1 t_1; \mathbf{k}_2\sigma_2 t_2 \right)$, where $t_c = \frac{1}{2}(t_1 + t_2)$ and $t_r = t_2 - t_1$ denote the center-of-mass and relative times, respectively.  
The corresponding center-of-mass frequency $\Omega_c$ and the inner-pair relative frequency $\omega_r$ in $\Gamma^{(2)}$ are defined as
\begin{equation}
\Omega_c = \frac{1}{\hbar} ( E_1^{(2)} + E_2^{(2)} ) , \  \omega_r = \frac{1}{2\hbar} ( E_2^{(2)} - E_1^{(2)} ) , \label{eqn3.1.4}
\end{equation}
where $E_1^{(2)}$ and $E_2^{(2)}$ are the transferred energies in the two photoelectric processes 
\begin{eqnarray}
&& E^{(2)}_1 = \varepsilon^{(d)}_{\mathbf{k}_1+\mathbf{q}_1 \sigma_1} + \Phi - \hbar \omega_{\mathbf{q}_1} , \notag \\
&&  E^{(2)}_2 = \varepsilon^{(d)}_{\mathbf{k}_2+\mathbf{q}_2 \sigma_2} + \Phi - \hbar \omega_{\mathbf{q}_2} . \label{eqn3.1.5} 
\end{eqnarray}
Here $\varepsilon^{(d)}_{\mathbf{k}+\mathbf{q} \sigma}$ are the photoelectron energies with momentum $\mathbf{k}+\mathbf{q}$ and spin $\sigma$, $\hbar \omega_{\mathbf{q}}$ are the photon energies with momentum $\mathbf{q}$, and $\Phi$ is the target-material work function.

It is noted that the Bethe-Salpeter wave function $\phi^{(2)}_{\alpha\beta}\left( \mathbf{k}_1\sigma_1 t_1; \mathbf{k}_2\sigma_2 t_2 \right)$ describes the dynamical annihilation of two electrons of the target material. The coincidence detection probability $\Gamma^{(2)}$, which probes the frequency-space wave function $\phi^{(2)}_{\alpha\beta}(\mathbf{k}_1\sigma_1, \mathbf{k}_2\sigma_2; \Omega_c, \omega_r)$, therefore directly reveals dynamical two-body correlations in the target electron system. Furthermore, this approach enables explicit resolution of both center-of-mass and inner-pair relative frequency dependencies inherent to these dynamical two-body correlations \citep{SuZhang2020}.

In the pairing channel for Cooper pairs, the inner-pair relative dynamics is governed by a time-retarded attractive interaction, such as the phonon-mediated pairing interaction in conventional metallic superconductors. Therefore, the inner-pair relative dynamics encoded in the Bethe-Salpeter wave function obtained from cARPES can shed light on the microscopic pairing interaction governing the formation of Cooper pairs. This makes cARPES a vital tool for uncovering the microscopic pairing mechanism of superconductivity \citep{SuZhang2020, DevereauxPRB2023, CaoSuArXiv2024NJP, DevereauxJPSJ2021, ShenZXRMP2021, AlexandradinataSciPost2025}. 
In cuprate superconductors, the microscopic origin of superconductivity has remained a long-standing puzzle. Does it arise from the coherent motion of holons with paired spinons, as suggested by the resonating valence bond (RVB) theory \citep{AndersonScience1987, PALeeRMP2006}? Or is it mediated by a pairing ``glue" such as dynamical spin fluctuations in a nearly antiferromagnetic Fermi liquid \citep{MonthouxPinesPRB1993, MonthouxPinesPRB1994}? Could it even be related to a Feshbach resonance scenario \citep{GrusdtNatCommu2023, GrusdtPRB2024, GrusdtNatCommu2025}? Measurements via cARPES in the pairing channel hold the potential to resolve this fundamental question. 
Moreover, cARPES is poised to make significant contributions to identifying the microscopic pairing mechanisms in other classes of unconventional superconductors, including iron-based superconductors \citep{ChenXHNAR2014, StewartFeSCRMP2011} and heavy-fermion superconductors \citep{Stewart1984, StewartNFLRMP2001}. It has also been proposed recently as a promising technique for probing the fluctuating nature of light-enhanced d-wave superconductivity \citep{ShiTaoPRX2021}, as well as the off-diagonal matrix-element driven spin spectra in topological states \citep{LinNatCommu2024}.

In the initial experimental design for cARPES measurement \citep{SuZhang2020,CaoSuArXiv2024NJP}, a coincidence detector is employed to instantaneously record the simultaneous occurrence of two photoelectric processes. This configuration is termed {\it instantaneous} cARPES. More recently, an alternative implementation was proposed (schematically shown in Fig.~\ref{fig1} (e)), termed  {\it post-experiment} cARPES \citep{CaoSuArXiv2024NJP}. To capture the dynamical two-body correlations in the target electron system, the time interval between the initiations of two photoelectric processes must be shorter than the characteristic time scale $t_c$ of the physics under investigation. This can be achieved using a pulse photon source, where both incident photons originate from the same pulse with a time width $\Delta t_p \leq t_c$. Furthermore, to clearly distinguish individual coincidence detection events, the time interval $\Delta t_d$ between consecutive photon pulses must satisfy $\Delta t_d \gg \Delta t_p, t_c$.
A further refinement in the  post-experiment cARPES is the use of two counting recorders $R_1$ and $R_2$, which record photoelectrons arriving at two detectors $D_1$ and $D_2$, respectively. For the $n$-th photon pulse, the recorded photoelectron numbers $a_1(n)$ and $a_2(n)$ in $R_1$ and $R_2$ take values of 0 or 1. The coincidence counting for the $n$-th pulse is then computed as $a_1(n)\times a_2(n)$ during post-processing. This approach, without the coincidence detector used in the instantaneous cARPES, enables easier and more efficient identification of two-photoelectric-process coincidence events. As a result, the post-experiment cARPES technique offers a highly efficient and powerful mean for measuring two-body correlations in the target  electron system \citep{CaoSuArXiv2024NJP}.

It is worth noting that the cARPES technique shares conceptual similarity with time-resolved correlation ARPES measurements of two-electron emission events \citep{StahlEcksteinNoiseARPESPRB2019}. Both techniques have the similar fundamental formulations based on perturbation S-matrix theory  \citep{SuZhang2020, DevereauxPRB2023, CaoSuArXiv2024NJP}. While cARPES measurements probe dynamical two-body correlations in the equilibrium states, time-resolved correlation ARPES measurements investigate these correlations in time-evolution nonequilibrium states. By incorporating the temporal envelope of ultrashort photon pulses, time-resolved correlation ARPES measurements enable time-resolved noise two-body correlations in nonequilibrium states. This contrasts with cARPES measurements, which provide dynamical two-body correlations in equilibrium states within frequency space, resolving both center-of-mass and relative frequency dependencies. More recently, a variant termed correlation ARPES technique has been developed from its time-resolved predecessor \citep{KemperPRB2025}. While it has the same theoretical formulation as the cARPES technique \citep{SuZhang2020, DevereauxPRB2023, CaoSuArXiv2024NJP}, this approach emphasizes the extraction of information on electron-electron interactions in weakly correlated electron systems through the coincidence detection of two photoelectric processes.

\subsection{{cINS}} \label{sec3.2}

\begin{techdefinition}
\noindent {\it Coincidence detection of two neutron-scattering processes is essentially a measurement of two-spin correlations in the target magnetic material}. --- \text{cINS}
\end{techdefinition}

A coincidence detection technique in the spin-spin channel, termed cINS, can be analogously proposed \citep{SucINS2021}. Here, the two incident particles are neutrons, which interact with the local spin moments of the target magnetic material and become scattered. These scattered neutrons can be coincidentally detected either by a dedicated coincidence detector \citep{SucINS2021} or via a post-experiment coincidence counting method \citep{CaoSuArXiv2024NJP}.

The two neutron-scattering processes are governed by the following electron-neutron interaction terms \citep{Lovesey1984,Squires1996,FelixPrice2013,SucINS2021,CaoSuArXiv2024NJP}
\begin{eqnarray}
V_1 &=& \sum_{\sigma_{i_1} \sigma_{f_1}} g(\mathbf{q}_1) f^{\dag}_{\mathbf{q}_{f_1} \sigma_{f_1}} \boldsymbol{\tau}_{\sigma_{f_1} \sigma_{i_1}} f_{\mathbf{q}_{i_1} \sigma_{i_1}} \cdot \mathbf{S}_{\perp}(\mathbf{q}_1) , \notag \\
V_2 &=& \sum_{\sigma_{i_2} \sigma_{f_2}} g(\mathbf{q}_2) f^{\dag}_{\mathbf{q}_{f_2} \sigma_{f_2}} \boldsymbol{\tau}_{\sigma_{f_2} \sigma_{i_2}} f_{\mathbf{q}_{i_2} \sigma_{i_2}} \cdot \mathbf{S}_{\perp}(\mathbf{q}_2) .
\label{eqn3.2.1}
\end{eqnarray}
Here, $f^{\dag}_{\mathbf{q} \sigma}$ and $f_{\mathbf{q} \sigma}$ denote the creation and annihilation operators for a neutron with momentum $\mathbf{q}$ and spin $\sigma$, while $\boldsymbol{\tau}$ represents the vector of Pauli matrices. The quantity $\mathbf{S}_{\perp}(\mathbf{q}) = \mathbf{S}(\mathbf{q}) \cdot (1 - \widehat{\mathbf{q}}\widehat{\mathbf{q}})$ is the transverse component of the spin operator, where $\mathbf{S}(\mathbf{q}) = \sum_{l} \mathbf{S}_l e^{-i\mathbf{q}\cdot \mathbf{R}_l}$ with $\mathbf{S}_l$ being the electron spin operator at site $\mathbf{R}_l$, and $\widehat{\mathbf{q}} = \mathbf{q}/|\mathbf{q}|$. In Eq. (\ref{eqn3.2.1}), $\mathbf{q}_1 = \mathbf{q}_{f_1}- \mathbf{q}_{i_1}$ and $\mathbf{q}_2 = \mathbf{q}_{f_2}- \mathbf{q}_{i_2}$ describe the momentum transfers in the two neutron-scattering processes.

Let us define a two-spin Bethe-Salpeter wave function as \citep{SucINS2021,CaoSuArXiv2024NJP}
\begin{equation}
\phi^{(ij)}_{\alpha\beta}(\mathbf{q}_1 t_1, \mathbf{q}_2 t_2) = \langle \Psi_\beta \vert T_t S^{(j)}_\perp (\mathbf{q}_2, t_2) S^{(i)}_\perp (\mathbf{q}_1, t_1) \vert \Psi_\alpha \rangle ,  \label{eqn3.2.2}
\end{equation}
where $i$ and $j$ label the Cartesian components of the spin operator. It can be shown that the coincidence detection probability $\Gamma^{(2)}$ for the simultaneous occurrence of two neutron-scattering processes provides direct information on $\phi^{(ij)}_{\alpha\beta}(\mathbf{q}_1 , \mathbf{q}_2; \Omega_c,\omega_r)$, which is obtained via Fourier transformation of $\phi^{(ij)}_{\alpha\beta}(\mathbf{q}_1, \mathbf{q}_2; t_c, t_r ) = \phi^{(ij)}_{\alpha\beta}(\mathbf{q}_1 t_1, \mathbf{q}_2 t_2)$.

Since the two-spin Bethe-Salpeter wave function $\phi^{(ij)}_{\alpha\beta}(\mathbf{q}_1 t_1, \mathbf{q}_2 t_2)$ describes the time evolution of the magnetic system with two spins excited at times $t_1$ and $t_2$, the coincidence detection probability effectively probes the dynamical two-spin correlations, resolving both center-of-mass and relative internal dynamics of the spin pairs. As a result, cINS emerges as a promising technique for investigating two-spin correlations in quantum magnetic materials \citep{SucINS2021}. It holds particular potential for uncovering new physics inaccessible to traditional INS measurements, including the long-sought quantum spin liquids in quantum magnets \citep{BalentsNature2010, ZhouYiRMP2017, ChamorroCR2021}.

\subsection{{cARP/IPES} and {cARIPES}} \label{sec3.3}

\begin{techdefinition}
\noindent {\it Coincidence detection of one photoelectric process and one inverse-photoemission process probes two-body correlations of the correlated electrons in the particle-hole channel}. --- \text{cARP/IPES} \\

\noindent {\it Coincidence detection of two inverse-photoemission processes accesses two-body correlations of the correlated electrons in the particle-particle channel}. --- \text{cARIPES}
\end{techdefinition}

The cARPES method can be extended to two other coincidence detection techniques, coincidence angle-resolved photoemission and inverse-photoemission spectroscopy (cARP/IPES) and coincidence angle-resolved inverse-photoemission spectroscopy (cARIPES).

In the cARP/IPES measurement, one photon source and one electron source are employed. The incident photon induces a photoelectric process, while the incident electron triggers an inverse-photoemission process, in which it transits into a low-energy state of the target material while simultaneously emitting a photon. The two relevant electron-photon interaction terms governing the photoelectric and inverse-photoemission processes are given by \citep{SuZhang2020}
\begin{eqnarray}
V_1 &=&  g_1 d^{\dag}_{\mathbf{k}_1+\mathbf{q}_1 \sigma_1} c_{\mathbf{k}_1\sigma_1} a_{\mathbf{q}_1\lambda_1} , \notag \\
V_2 &=&  g_2 c^{\dag}_{\mathbf{k}_2\sigma_2} a^{\dag}_{\mathbf{q}_2\lambda_2} d_{\mathbf{k}_2+\mathbf{q}_2 \sigma_2}  . \label{eqn3.3.1}
\end{eqnarray}
Here, the creation and annihilation operators and the coupling constants $g_1$ and $g_2$ are defined as in Eq.~(\ref{eqn3.1.1}).

Coincidence detection of the photoelectron emitted in the photoelectric process and the photon emitted in the inverse-photoemission process yields the joint probability of their simultaneous occurrence. The corresponding coincidence detection probability captures the two-body correlations in the particle-hole channel, which can be described by a two-body Bethe-Salpeter wave function defined as \citep{SuZhang2020}
\begin{equation}
\phi^{(3)}_{\alpha\beta}\left( \mathbf{k}_1\sigma_1 t_1; \mathbf{k}_2\sigma_2 t_2 \right) = \langle \Psi_{\beta} \arrowvert T_t c^{\dag}_{\mathbf{k}_2 \sigma_2} (t_2) c_{\mathbf{k}_1 \sigma_1} (t_1) \arrowvert \Psi_{\alpha} \rangle . \label{eqn3.3.2}
\end{equation}
More specifically, cARP/IPES provides direct access to its Fourier transformation $\phi^{(3)}_{\alpha\beta}\left(\mathbf{k}_1\sigma_1, \mathbf{k}_2\sigma_2; \Omega_c, \omega_r \right)$, where $\Omega_c$ and $\omega_r$ denote the center-of-mass and inner-pair relative frequencies associated with the particle-hole pairs excited in the relevant photoelectric and inverse-photoemission processes.

cARP/IPES is poised to serve as a powerful tool for elucidating the long-standing puzzle of the microscopic mechanism underlying itinerant magnetism in metallic ferro- and antiferromagnets \citep{Moriya1985}. In such systems, the magnetic moments originate fundamentally from particle-hole correlations in the spin channel. Moreover, cARP/IPES is expected to contribute significantly to the study of metallic nematic states \citep{FradkinARCMP2010,SuLi2015,SuLi2017}, which are governed by particle-hole correlations in the charge channel. It has also been proposed as a promising technique for observing collective Fermi arcs arising from particle-hole excitations in topological Weyl semimetals \citep{TzengWeylPRB2020}.

In the cARIPES technique, two electron sources are utilized. Electrons from these sources initiate two inverse-photoemission processes, respectively, during which they are injected into the target material, transition to its low-energy electronic states, and simultaneously emit two photons into the vacuum. The electron-photon interaction terms governing these two inverse-photoemission processes are given by
\begin{eqnarray}
V_1 &=&  g_1 c^{\dag}_{\mathbf{k}_1\sigma_1} a^{\dag}_{\mathbf{q}_1\lambda_1}  d_{\mathbf{k}_1+\mathbf{q}_1 \sigma_1} , \notag \\
V_2 &=&  g_2 c^{\dag}_{\mathbf{k}_2 \sigma_2} a^{\dag}_{\mathbf{q}_2\lambda_2} d_{\mathbf{k}_2+\mathbf{q}_2 \sigma_2} . \label{eqn3.3.3}
\end{eqnarray}

Coincidence detection of the two emitted photons reveals the joint probability of the simultaneous occurrence of the two inverse-photoemission processes, which is described by a two-body Bethe-Salpeter wave function in the particle-particle channel \citep{SuZhang2020}
\begin{equation}
\phi^{(4)}_{\alpha\beta}\left( \mathbf{k}_1\sigma_1 t_1; \mathbf{k}_2\sigma_2 t_2 \right) = \langle \Psi_{\beta} \arrowvert T_t c^{\dag}_{\mathbf{k}_2 \sigma_2} (t_2) c^{\dag}_{\mathbf{k}_1 \sigma_1} (t_1) \arrowvert \Psi_{\alpha} \rangle . \label{eqn3.3.4}
\end{equation}
Physically, this Bethe-Salpeter wave function captures the dynamical correlated creation of two electrons within the material, thereby characterizing the dynamical correlations of particle-particle pairs. More specifically, the coincidence detection probability in cARIPES directly yields the Fourier-transformed Bethe-Salpeter wave function $\phi^{(4)}_{\alpha\beta}(\mathbf{k}_1\sigma_1, \mathbf{k}_2\sigma_2; \Omega_c, \omega_r)$, which distinctly resolves both center-of-mass and inner-pair relative dynamical two-body correlations. Unlike the cARPES case, the excited electrons described by $\phi^{(4)}_{\alpha\beta}(\mathbf{k}_1\sigma_1 t_1; \mathbf{k}_2\sigma_2 t_2)$ reside predominantly above the Fermi energy. As a result, cARIPES probes two-body correlations in the particle-particle channel mainly for the electrons above the Fermi energy, making it uniquely suited for investigating Cooper-pair physics in which the paired electrons primarily occupy states above the Fermi energy \citep{SuZhang2020}.

\subsection{{Coincidence double-tip STS}} \label{sec3.4}

\begin{techdefinition}
\noindent {\it Coincidence detection of two quantum tunneling currents through two local tips reveals the spatially resolved dynamical two-body correlations of the correlated electrons}. --- \text{Coincidence double-tip STS}
\end{techdefinition}

A coincidence detection technique for directly probing spatially resolved dynamical two-body correlations has recently been proposed, based on the double-tip scanning tunneling microscope (STM) and termed coincidence double-tip STS \citep{SuArXiv2025}. A schematic illustration of its setup is provided in Fig.~\ref{fig1} (d).

The coincidence double-tip STS setup involves two probe tips located at positions $j_1$ and $j_2$, biased at distinct voltages $V_1$ and $V_2$ respectively, with each tip carrying a quantum tunneling current. At time $t$, the measured tunneling currents are denoted as $I_1(t,n)$ and $I_2(t,n)$, where $n$ indexes different measurements. A coincidence tunneling current observable $I_d^{(2)}$ is defined as
\begin{equation}
I_d^{(2)} = \frac{1}{N_n} \sum_{n=1}^{N_n} I_1(t, n) \cdot I_2(t, n) , \label{eqn3.4.1}
\end{equation}
with $N_n$ being the total number of measurements. Theoretically, $I_d^{(2)}$ corresponds to the following statistically averaged correlation function of two local tunneling current operators $I_1$ and $I_2$ at positions $j_1$ and $j_2$, respectively, defined by 
\begin{equation}
I_d^{(2)}=\overline{\langle I_1(t) I_2(t)\rangle} . \label{eqn3.4.2}
\end{equation}
By introducing a coincidence dynamical conductance 
\begin{equation}
\sigma_{j_1 j_2}(V_1, V_2) = \frac{d}{d V_2} \frac{d}{d V_1} \overline{\langle I_1(t) I_2(t)\rangle} , \label{eqn3.4.3}
\end{equation}
a nonequilibrium theoretical framework yields the following expression 
\begin{eqnarray}
\sigma_{j_1 j_2}(V_1, V_2) &=& -\left(\frac{e^2 T_1 T_2}{2\pi \hbar^2}\right)^2 \sum_{\sigma_1 \sigma_2} G^{(2)}_{J}(j_1\sigma_1, j_2\sigma_2; \omega_1, \omega_2 ) \notag \\
&& \times A_{d_1}(0) A_{d_2}(0). \label{eqn3.4.4}
\end{eqnarray}
Here, $G^{(2)}_{J}$ denotes a second-order contour-time ordered current correlation function defined by
\begin{eqnarray}
&&G^{(2)}_{J}(j_1\sigma_1,j_2\sigma_2; \omega_1, \omega_2) \notag \\
&=& \iint_c d t_2 d t_1 \langle T_c J_{1A} (j_1\sigma_1; t, t_1; \omega_1) J_{2A} (j_2\sigma_2; t, t_2; \omega_2) \rangle , \notag \\
&& \label{eqn3.4.5}
\end{eqnarray}
where $T_c$ represents the contour-time ordering operator defined along a closed time contour $t_i \rightarrow t \rightarrow t_i$, and the integrations are performed along this time contour. The effective current operators $J_{1A}$ and $J_{2A}$ are given by
\begin{eqnarray}
&& J_{1A}(j_1\sigma_1; t, t_1; \omega_1) = -i c_{j_1\sigma_1}(t) c^\dag_{j_1\sigma_1} (t_1) e^{-i\omega_1 (t_1 - t)} + h.c. , \notag \\
&& J_{2A}(j_2\sigma_2; t, t_2; \omega_2) = -i c_{j_2\sigma_2}(t) c^\dag_{j_2\sigma_2} (t_2) e^{-i\omega_2 (t_2 - t)} + h.c. . \notag \\ 
&& \label{eqn3.4.6}
\end{eqnarray}
The frequencies $\omega_1$ and $\omega_2$ are related to the bias voltages by $\omega_1 = -e V_1/ \hbar$ and $\omega_2 = - e V_2 /\hbar$. In Eq.~(\ref{eqn3.4.4}), $T_1$ and $T_2$ are the tunneling constants for the two tunneling currents, while $A_{d_1}(0)$ and $A_{d_2}(0)$ represent the single-particle spectral functions of the two tip-electron systems at their Fermi energies.

The coincidence dynamical conductance $\sigma_{j_1 j_2}(V_1, V_2)$ provides access to dynamical two-body correlations that go beyond the physics captured by traditional single-tip STS technique \citep{SuArXiv2025}. For instance, in a nearly free Fermi liquid, the coincidence dynamical conductance reveals two correlated dynamical electron propagation processes: (i) from $j_1$ to $j_2$ (or vice versa) driven by $V_1$, and (ii) from $j_2$ to $j_1$ (or vice versa) driven by $V_2$. In a superconducting state, additional propagation channels originating from the superconducting condensate emerge, coexisting with the normal electron propagation processes.

Experimental implementation of the double-tip STS has advanced considerably over the past two decades \citep{SHIRAKI2001633, NakayamaAdM201200257, BertRevSI2018, LeeuwenhoekPRB2020, MaartenNature2020}. However, achieving precise control over tip positionings at extremely short length scales remains a major challenge. Modern double-tip STM setups have successfully realized tip separations on the order of $30~\text{nm}$ \citep{NakayamaAdM201200257, MaartenNature2020}. Nevertheless, such resolutions are still insufficient for studying most strongly correlated electron systems, such as various classes of unconventional superconductors \citep{PALeeRMP2006, ChenXHNAR2014, StewartFeSCRMP2011}, where dominant strong correlations are highly localized, with typical correlation lengths comparable to interatomic distances (a few angstroms).
The recent discovery of unconventional superconductivity and Mott physics in magic-angle twisted bilayer graphene \citep{CaoNature2018} offers a promising alternative platform for coincidence double-tip STS. In this system, exotic correlated phenomena are inherently connected to the moir\'{e} superlattice, which exhibits a characteristic periodicity of approximately $13 ~\text{nm}$, a mesoscopic length scale within the resolution of existing double-tip STM setups. Consequently, moir\'{e} quantum materials represent an ideal platform for implementing coincidence double-tip STS to probe spatially resolved dynamical two-body correlations in strongly correlated electron systems.

\subsection{Double Photoemission} \label{sec3.5}

\begin{techdefinition}
\noindent {\it The one-photon-in and two-electron-out scattering processes, with the two emitted electrons coincidentally detected, provide information on the center-of-mass physics of the two-body correlations in the particle-particle channel}. --- \text{Double photoemission}
\end{techdefinition}

In 1998, Berakdar formulated the $(\gamma, 2e)$ photoemission technique for studying the correlated electrons in solids, now commonly known as double photoemission \citep{BerakdarPRB1998}. This theoretical framework was subsequently extended to investigate the correlated electrons on solid surfaces \citep{FominykhPRL2002} and in various superconducting pairing states \citep{BerakdarSCPRL2003,MorrPRB2022}.

In the double photoemission measurement, a single incident photon excites two correlated electrons from the target material, constituting a one-photon-in and two-electrons-out scattering event. Since the fundamental electrodynamics dictates that a single photon can typically excite only one electron in a photoelectric process, double photoemission must involve additional electron-electron interactions. The relevant interaction terms in the double photoemission measurement comprise the electron-photon interaction $V_1$ and the electron-electron interaction $V_2$, defined respectively as
\begin{equation}
V_1(t) = \sum_{\mathbf{k} \mathbf{q}\lambda\sigma} g \, d^\dag_{\mathbf{k}+\mathbf{q} \sigma} (t) c_{\mathbf{k}\sigma} (t) a_{\mathbf{q}\lambda} (t)  \label{eqn3.5.1}
\end{equation}
and
\begin{eqnarray}
V_{2}(t_1,t_2) &=& \frac{1}{2} \sum_{\mathbf{k}_1 \mathbf{k}_2 \mathbf{q}_1 \sigma_1 \sigma_2} U f^\dag_{\mathbf{k}_1+\mathbf{q}_1 \sigma_1}(t_1) e_{\mathbf{k}_1 \sigma_1} (t_1) \notag \\
&& \times d^\dag_{\mathbf{k}_2 -\mathbf{q}_1 \sigma_2}(t_2)  c_{\mathbf{k}_2 \sigma_2 }(t_2) ,  \label{eqn3.5.2}
\end{eqnarray}
where the operators $c, d, e, f$ in $V_2$ represent the electrons in possibly different energy bands. Here, the interaction constants $g\equiv g(\mathbf{k};\mathbf{q}\lambda)$ and $U\equiv U(\mathbf{k}_1, \mathbf{k}_2, \mathbf{q}_1; t_1, t_2)$. It is noted that $U$ generally exhibits strong time-dependent dynamics, which may stem from electron-phonon interactions or spin/charge-density wave fluctuations. For brevity, the band-index dependence of $g$ and $U$ is not explicitly shown.

It is crucial to note that $V_2$ does not originate from an externally applied field but arises from the intrinsic electron-electron interactions within the material. Consequently, it can not be directly tuned experimentally. As a result, double photoemission signals are inherently entangled with complex Coulomb drag processes \citep{DevereauxPRB2023}. Although the coincidence detection probability of the double photoemission measurement contains information on the two-body Bethe-Salpeter wave function in the particle-particle channel, it also includes convoluted contributions from summation over inner-pair momenta and/or frequency integration \citep{MorrPRB2022,SuYHJPCM2024}. This entanglement precludes clear resolution of the inner-pair spatial and dynamical physics of the two-body correlations in the target electron system.

A simple explanation for the inability to resolve the inner-pair physics can be given as follows. Suppose the incident photon carries momentum $\mathbf{q}$ and energy $\hbar \omega _{\mathbf{q}}$, while the two emitted photoelectrons have momenta ($\mathbf{k} _{1}^\prime, \mathbf{k} _{2}^\prime$) and energies ($\varepsilon_{\mathbf{k}^\prime_{1}},\varepsilon_{\mathbf{k}^\prime_{2}}$). Momentum and energy conservations then yield
\begin{eqnarray}
&& \mathbf{q} = \mathbf{k} _{1}^\prime + \mathbf{k} _{2}^\prime - \Delta \mathbf{k}, \,\,  \Delta \mathbf{k} = \mathbf{k} _{1} + \mathbf{k} _{2}, \notag \\
&& \hbar \omega_\mathbf{q} =  \varepsilon_{\mathbf{k}^\prime_{1}} + \varepsilon_{\mathbf{k}^\prime_{2}} - \Delta E, \,\, \Delta E = E_1 + E_2 , \label{eqn3.5.3} 
\end{eqnarray}
where ($\mathbf{k}_{1},\mathbf{k}_{2}$) and ($E_{1},E_{2}$) denote the momenta and energies of the two correlated electrons inside the target material. Therefore, in the double photoemission measurement with the fixed parameters $(\mathbf{q}, \mathbf{k} _{1}^\prime, \mathbf{k} _{2}^\prime, \hbar \omega_\mathbf{q}, \varepsilon_{\mathbf{k}^\prime_{1}} , \varepsilon_{\mathbf{k}^\prime_{1}})$, the sum of the momenta $\mathbf{k}_{1}$ and $\mathbf{k}_{2}$ and the sum of the energies $E_{1}$ and $E_{2}$ are determined, but the corresponding differences remain unconstrained. This is the fundamental reason why double photoemission can not resolve the internal structure of the two-body correlations in the target electron system.

Nevertheless, since the center-of-mass momentum and energy of the two-body correlations can be accessed via Eq. (\ref{eqn3.5.3}), double photoemission remains a valuable technique for probing the center-of-mass physics of correlated electron pairs. For instance, Mahmood {\it et al}. provided a detailed theoretical study of coincidence two-body spectral functions in the pair-density-wave (PDW) and Fulde-Ferrell-Larkin-Ovchinnikov (FFLO) phases, where Cooper pairs possess finite center-of-mass momentum \citep{MorrPRB2022}. Double photoemission can also be employed to investigate the collective modes of superconducting condensates \citep{SuYHJPCM2024}, such as the Goldstone-Anderson phase modes \citep{AndersonPlasmon1958}, the Higgs amplitude modes \citep{VarmaHiggsPRL1981,VarmaHiggsPRB1982}, and the Leggett modes in superconductors with two coupled macroscopic superconducting condensates \citep{LeggettMode1966}.

\section{Discussion and Future Prospects} \label{sec4}

In the preceding sections, we have briefly explored several recently proposed coincidence detection techniques in different measurement channels. These emerging methods show significant potential for elucidating the breakdown of Wick’s theorem and uncovering interaction-driven many-body correlations in quantum materials \citep{ShenZXRMP2021, AlexandradinataSciPost2025}.
With the exception of double photoemission, which has been under development for more than two decades, most coincidence detection techniques remain in their infancy. Future advances are anticipated along two primary directions, technological innovation and theoretical development.

On the technological front, a key challenge lies in the limited resolution of current experimental platforms. Addressing this challenge will require continuous, state-of-the-art improvements in both particle sources and detection systems \citep{ShenZXRMP2021, AlexandradinataSciPost2025}. For example, enhancing the resolution of coincidence measurements will depend on progress in the controlled emission and/or detection of photons, electrons, or neutrons with high precision in temporal, spatial, energy, and momentum domains. Insights drawn from well-established multiphoton coincidence detection in quantum optics \citep{ShihPRL1995Optics, PittmanPRA1995Optics, FerriPRL2005Optics, PanJW2012OpticsRMP} may offer valuable guidance in this context. Moreover, recent rapid advances in attosecond technology \citep{Gaumnitz:17, KimNature2023, LiNatCommu2017, Nobel2023} are expected to provide superior photon sources, which will substantially benefit studies of strongly correlated electron systems.

On the theoretical front, a central task is the accurate computation of coincidence detection probabilities, which are directly connected to dynamical two-body correlations in different physical channels. For instance, calculating coincidence detection probabilities for cARPES in the particle-particle pairing channel would provide essential theoretical support for upcoming experiments to study unconventional superconductors. Such calculations could help clarify the microscopic mechanisms of high-Tc superconductivity in cuprate, iron-based, and heavy-fermion superconductors \citep{AndersonScience1987, PALeeRMP2006, ChenXHNAR2014, StewartFeSCRMP2011, Stewart1984, StewartNFLRMP2001, MonthouxPinesPRB1993, MonthouxPinesPRB1994, GrusdtNatCommu2023, GrusdtPRB2024, GrusdtNatCommu2025}. Similarly, theoretical studies of coincidence detection probabilities for cINS would elucidate dynamical two-spin correlations in quantum magnets, potentially revealing new physics linked to quantum spin liquids \citep{BalentsNature2010, ZhouYiRMP2017}. Theoretical predictions for cARP/IPES coincidence detection in the particle-hole channel will be essential for understanding itinerant magnetism in the spin channel \citep{Moriya1985} and electronic nematicity in the charge channel \citep{FradkinARCMP2010}. Such efforts may further clarify the roles of antiferromagnetic and multiorbital fluctuations in the emergence of itinerant magnetism and d-wave nematic order in iron-based superconductors \citep{SuLi2015, SuLi2017}.

It is important to emphasize that the computation of coincidence detection probabilities in different dynamical two-particle channels remains highly challenging, both in theoretical analytical methods and numerical simulations. These challenges originate from the strong many-body correlations in strongly correlated electron systems, where a vast number of degrees of freedom are intrinsically entangled, as well as from the inherent complexity of dynamical two-body response functions. Consequently, new conceptual and methodological advances in theoretical analytical methods and numerical simulations will be essential to unravel the complex and profound physics encoded in coincidence detection probabilities.

Looking ahead, the development of novel coincidence detection techniques specifically tailored for strongly correlated electron systems holds great promise. Moreover, given the strong entanglement among spin, charge, and orbital degrees of freedom in such materials, approaches capable of directly probing correlations between these different degrees of freedom could open entirely new research directions. In particular, the coincidence detection of two physical processes involving distinct degrees of freedom would allow direct measurement of their mutual correlations and entanglement. Such methods hold the potential to greatly deepen our understanding of strongly correlated quantum matters, especially in shedding light on the complex intertwined quantum phases arising from such strong entanglement.

\section*{Acknowledgements}
We thank Prof. Jun Chang for invaluable discussions. This work was supported by the National Natural Science Foundation of China (Grant No. 11874318) and the Natural Science Foundation of Shandong Province (Grant No. ZR2023MA015). 





%

\end{document}